\documentclass{ws-rv975x65}
\usepackage{ws-rv-van}             

\makeindex

\begin{document}

\chapter{Heavy quark skyrmions\label{heavy skyrmions}}

\author[N.N. Scoccola]{N.N. Scoccola}

\address{Departmento de F\'\i sica, Comisi\'on Nacional de Energ\'{\i}a At\'omica, \\
(1429) Buenos Aires, Argentina.\\
CONICET, Rivadavia 1917, (1033) Buenos Aires, Argentina.\\
Universidad Favaloro, Sol{\'\i}s 453, (1078) Buenos Aires, Argentina.}

\begin{abstract}
The description of the heavy baryons as heavy-meson--soliton bound systems
is reviewed. We outline how such bound systems arise from
effective lagrangians that respect both chiral symmetry and heavy
quark symmetry. Effects due to finite heavy quark masses are also
discussed, and the resulting heavy baryon spectra are compared
with existing quark model and empirical results. Finally, we address
some issues related to a possible connection between the usual bound
state approach to strange hyperons and that for heavier baryons.
\end{abstract}

\body

\section{Introduction}
\label{intro}

During the last quarter of a century it has become clear that the
applicability of the Skyrme's topological soliton model for light
baryon structure\cite{Sky62,Zahed:1986qz} goes far beyond all the
original expectations. In fact, as described in other chapters of
this book the underlying ideas have found applications in other
areas of physics, notably in the physics of complex nuclei and
dense matter, condensed matter physics and gauge/string duality.
The purpose of the present contribution is to summarize the work
done on the extension of the skyrmion picture to the study of the
structure of baryons containing heavy quarks. In this scheme, such
baryons are described as bound systems of heavy mesons and a
soliton. This so-called "bound state approach" was first developed
to describe strange hyperons\cite{CK85,SNNR} and was later
shown\cite{RRS90} to be applicable to baryons containing one or
more charm (c) and bottom (b) quarks. In these early works only
pseudoscalar fields were taken as explicit degrees of freedom with
their interactions given by a flavor symmetric Skyrme lagrangian
supplemented by explicit flavor symmetric terms to account for the
effect of the heavy quark mass. The results on the mass
spectra\cite{RS91} and magnetic moments\cite{OMRS91} for charm
baryons were found to be strikingly close to the quark model
description which is expected to work better as the heavy quark
involved becomes heavier. However, it was then realized that this
description in terms of only pseudoscalar fields was at odds with
the heavy quark symmetry\cite{Manohar:2000dt} which states that in
the heavy quark limit the heavy pseudoscalar and vector fields
become degenerate and, thus, should be treated on an equal
footing. This difficulty was resolved in Ref.\cite{Jenkins:1992zx}
where it was proposed to apply the bound state approach to the
heavy meson effective lagrangian\cite{Wis92,BD92,Yan92,Goity:1992tp} which
simultaneously incorporates chiral symmetry and heavy quark
symmetry. Such observation led to a quite important number of
works where various properties of heavy baryons have been studied
within this framework. Here, we present a short review of those
studies pointing out their main results as well as the
relationship between some different approaches used in the
literature. Some still remaining open questions are also
mentioned.

This contribution is organized as follows. In Sec.\ref{heavybar}
we outline how heavy baryons can be described within soliton
models in the heavy quark limit. In particular, in
Subsec.\ref{effact} we introduce the type of lagrangian that
describes the interactions between light and heavy mesons, and
which simultaneously respect chiral and heavy quark symmetries,
while in Subsec.\ref{ubsa} we show how bound states of a soliton
and heavy mesons are obtained and the system quantized. In
Sec.\ref{beyhql} we show how departures from the heavy quark
limit can be taken into account. In Sec.\ref{relbsa} we discuss
some issues related to the connection between the usual bound state
approach to strange hyperons with that for heavier baryons given
in the previous section. Finally, in Sec.\ref{summary} a summary
with some conclusions is given.

\section{Heavy Baryons as Skyrmions in the Heavy Quark Limit}
\label{heavybar}

In this section we outline how a heavy baryon can be described
within topological soliton models in the limit in which the
heavy quarks are assumed to be infinitely heavy.
Corrections due to finite heavy quark masses will be
discussed in the following section. In Subsec.\ref{effact} we
introduce a type of lagrangian for a system of Goldstone bosons
and the heavy mesons, which possesses both chiral symmetry and
heavy quark symmetry. Next, in Subsec.\ref{ubsa} we show how a
heavy-meson--soliton bound state can arise at the classical level,
and the way in which such bound system can be quantized.

\subsection{Effective chiral lagrangians and heavy quark symmetry}
\label{effact}

For the light sector, the simplest lagrangian that supports stable
soliton configuration is the Skyrme model lagrangian \cite{Sky62}
\begin{equation}
{\cal L}_{l}^{Sk} = \frac{f^2_\pi}{4}
{\,\mbox{Tr}}\left[\partial_\mu U^\dagger \partial^\mu U \right] +
\frac{1}{32e^2} {\,\mbox{Tr}}\left[ [U^\dagger
\partial_\mu U,
   U^\dagger \partial_\nu U ]^2\right] ,
\label{Lsk}
\end{equation}
where $f_\pi$ is the pion decay constant
($\approx 93$ MeV empirically) and $U$ is an $SU(2)$ matrix of the
chiral field, i.e.
\begin{equation}
U = \exp\left[iM/f_\pi\right],
\end{equation}
with $M$ being a $2\times 2$ matrix of the pion triplet
\begin{equation}
M = \vec{\tau}\cdot\vec{\pi} =  \left( \!\! \begin{array}{cc}
\pi^0  \!\!&\!\! \sqrt2\pi^+ \\
\sqrt2\pi^- \!\!&\!\! -\pi^0  \end{array} \!\!\right).
\end{equation}
Here, the chiral $SU(2)_L\times SU(2)_R$ symmetry is realized
nonlinearly under the transformation of $U$
\begin{equation}
U \longrightarrow L\ U\ R^\dagger,
\end{equation}
with $L \in SU(2)_L$ and $R \in SU(2)_R$. Due to the presence of
the Skyrme term with the Skyrme parameter $e$, the lagrangian
${\cal L}_l^{Sk}$ supports stable soliton solutions.

When discussing the interaction of the Goldstone fields $M(x)$
with other fields it is convenient to introduce $\xi(x)$ such that
\begin{equation}
U = \xi^2,
\label{xi}
\end{equation}
and which transforms under the $SU(2)_L\times SU(2)_R$ as
\begin{equation}
\xi \rightarrow \xi^\prime = L \ \xi  \ \vartheta^\dagger
= \vartheta \ \xi \ R^\dagger,
\label{Xct}
\end{equation}
where $\vartheta$ is a local unitary matrix depending
on $L$, $R$, and $M(x)$.

Consider now heavy mesons containing a heavy quark $Q$ and a light
antiquark $\bar{q}$. Here, the light antiquark in a heavy meson is
assumed to form a point-like object with the heavy quark, endowing
it with appropriate color, flavor, angular momentum and parity.
Let $\Phi$ and $\Phi_\mu^*$ be the field operators that annihilate
$j^\pi$=$0^-$ and $1^-$ heavy mesons with $C=+1$ or $B=-1$. They
form $SU(2)$ antidoublets: for example, when the heavy
quark constituent is the $c$-quark,
\begin{equation}
\Phi = (D^0, D^+)  \qquad , \qquad
\Phi^* = (D^{*0}, D^{*+})\ .
\label{Phi}
\end{equation}

In the limit of infinite heavy quark mass, the heavy quark symmetry
implies that the dynamics of the heavy mesons depends trivially on
their spin and mass. Such a trivial dependence can be eliminated by
introducing a redefined $4\times4$ matrix field $H(x)$
as
\begin{equation}
H = \frac{1+{v \!\!\!/}}{2} \left( \Phi_v \gamma_5
   - \Phi^*_{v\mu}\gamma^\mu \right).
\label{H}
\end{equation}
Here, we use the conventional Dirac $\gamma$-matrices and
${v \!\!\!/}$ denotes $v_\mu\gamma^\mu$. The fields $\Phi_v$ and
$\Phi_{v\mu}^*$, respectively, represent the heavy pseudoscalar
field and heavy vector fields in the moving frame with a four
velocity $v_\mu$. They are related to the $\Phi$ and $\Phi^*_\mu$
as\cite{Geo92}
\begin{equation}
\Phi = e^{-iv\cdot x m_\Phi} \frac{\Phi_v}{\sqrt{2m_\Phi}}
\quad , \quad
\Phi^*_\mu = e^{-iv\cdot x m_{\Phi^*}}
\frac{ \Phi^*_{v\mu}}{\sqrt{2m_{\Phi^*}}}\ .
\end{equation}
Under $SU(2)_L\times SU(2)_R$ chiral symmetry operations
$H$ transforms as
\begin{equation}
H \rightarrow H \ \vartheta\ ,
\end{equation}
while under the heavy quark spin rotation,
\begin{equation}
H \rightarrow S\ H\ ,
\label{hb}
\end{equation}
with $S\in SU(2)_v$, i.e. the heavy quark spin symmetry group boosted by
the velocity $v$. Taking this into account it is possible to write down
a lagrangian that describes the interactions of heavy mesons and Goldstone
bosons, and which possesses both chiral symmetry and heavy quark
symmetry. To leading order in derivatives acting on the Goldstone fields,
the most general form of such lagrangian is given by\cite{Wis92,BD92,Yan92,Goity:1992tp}
\begin{equation}
{\cal L}_{lh} = - iv_\mu{\,\mbox{Tr}}\left[D^\mu H\bar{H}\right]
 - g{\,\mbox{Tr}}(\left[H\gamma_5 A_\mu\gamma^\mu\bar{H}\right]\ ,
\label{LvHx}
\end{equation}
where $\bar{H}=\gamma_0 H^\dagger\gamma_0,$ and
\begin{equation}
V_\mu=\frac12(\xi^\dagger\partial_\mu\xi
             +\xi\partial_\mu\xi^\dagger)
\qquad , \qquad
A_\mu=\frac{i}2(\xi^\dagger\partial_\mu\xi
             -\xi\partial_\mu\xi^\dagger)\ .
\label{currents}
\end{equation}
Here, $g$ is a universal coupling constant  for the $\Phi\Phi^*\pi$
and $\Phi^*\Phi^*\pi$ interactions. The nonrelativistic quark model
provides the naive estimation\cite{Yan92} $g=-3/4$.
On the other hand, for the case of the  $D^* \rightarrow \pi D$ decay
this lagrangian leads to a width given by
\begin{equation}
\Gamma(D^{*+} \rightarrow D^0 \pi^+ )
= \frac{1}{6\pi}\frac{g^2}{f_\pi^2}|\vec{p}_\pi|^3\ .
\end{equation}
Recent experimental results for this width
imply $|g|^2 \approx 0.36$ \cite{Anastassov:2001cw}.

\subsection{Heavy-Meson--Soliton Bound States in the Heavy Quark Limit and their
Collective Quantization}
\label{ubsa}

Following the discussions in the previous subsection we consider
here the chiral and heavy quark symmetric effective lagrangian
given by
\begin{equation}
{\cal L} = {\cal L}^{Sk}_{l} + {\cal L}_{lh} \ ,
\label{leff}
\end{equation}
where ${\cal L}^{Sk}_{l}$ and ${\cal L}_{lh}$ are given by Eqs.(\ref{Lsk}) and (\ref{LvHx}),
respectively.

In what follows we will discuss how to obtain heavy baryons
following a procedure in which a heavy-meson--soliton bound state
is first found and then quantized by rotating the whole
system in the collective coordinate quantization
scheme\cite{Gupta:1993kd,Min:1994qq}. An alternative
method\cite{Jenkins:1992zx} will be briefly discussed at the end
of this subsection.

The non-linear lagrangian ${\cal L}^{Sk}_l$ supports a classical
soliton solution
\begin{equation}
U_0(\vec r) = \exp[i\vec\tau\cdot\hat{r}F(r)]\ ,
\label{CSC}
\end{equation}
with the boundary conditions
\begin{equation}
F(0)=\pi \qquad \mbox{and} \qquad F(\infty)=0 \ ,
\label{CSCa}
\end{equation}
which, due to its nontrivial topological structure, carries a
winding number identified as the baryon number $B=1$. It also has
a finite mass $M_{sol}$ whose explicit expression in terms of the
soliton profile function $F(r)$ can be found in e.g.
Refs.\cite{Sky62,Zahed:1986qz} .

In order to look for possible heavy-meson--soliton bound states we have
to find the eigenstates of the heavy meson fields interacting with the static
potentials
\begin{eqnarray}
V^\mu &=& \left( 0, \vec V \right)
= \Big(0,i\ v(r) \ \hat{r} \times\vec \tau \Big) \ , \nonumber \\
A^\mu &=& \left( 0, \vec A \right) = \Big( 0, \frac12\ a_1(r) \
\vec{\tau} + \frac12\ a_2(r) \ \hat{r}\ \vec \tau \cdot \hat{r}
\Big), \label{AV}
\end{eqnarray}
where
\begin{eqnarray}
v(r) = \frac{\sin^2(F/2)}{r} \!\!\!\! \qquad , \qquad \!\!\!\!
a_1(r) = \frac{\sin\!F}{r}  \!\!\!\! \qquad , \qquad \!\!\!\!
a_2(r) = F^\prime - \frac{\sin\!F}{r}\ .
\end{eqnarray}
These expressions result from the soliton configuration (\ref{CSC}) sitting at the
origin. In the rest frame $v_\mu = (1,0,0,0)$, it follows from Eq.(\ref{H}) that
$H(x)$ can be expressed in terms of 2 x 2 blocks as
\begin{equation}
H(x) = \left(%
\begin{array}{cc}
  0 & h(x) \\
  0 & 0 \\
\end{array}%
\right)\ .
\end{equation}
Here we have used that, in that frame, $\Phi^*_{v,0}$ is identically zero due to the
condition $v \cdot \Phi^*_{v} =0$.
Thus, the lagrangian Eq.(\ref{LvHx}) takes the form
\begin{equation}
L_0 = -M_{sol} +
\int\!\!d^3r\left( -i{\,\mbox{Tr}}\left[ \dot h \ \bar{h} \right] +
g {\,\mbox{Tr}}\left[ h \ \vec A\cdot\vec\sigma \ \bar{h} \right] \right)\ ,
\label{L0msb}
\end{equation}
where $\bar h = - h^\dagger$. The corresponding equation
of motion for the $h$-field is\cite{Oh:1994yv,Min:1994qq}
\begin{equation}
i \ \dot h = g \ h\ \vec{A}\cdot{\vec\sigma}\ . \label{hveq}
\end{equation}
In the ``hedgehog" configuration (\ref{CSC}), and consequently
in the static potentials (\ref{AV}), the isospin and the angular
momentum are correlated in such a way that neither of them is
separately a good quantum number, but their sum (the so-called
"grand spin") $\vec K$ is. Here
\begin{equation}
\vec K=\vec J+\vec I\equiv (\vec L+\vec S)+\vec I\ .
\end{equation}
Thus, the equation of motion Eq.(\ref{hveq}) is invariant under
rotations in $K$-space, and the wavefunctions of the heavy meson
eigenmodes can be written as the product of a radial function
and the eigenfunction of the grand spin ${\cal K}^{(a)}_{kk_3}(\hat
r)$. Namely,
\begin{equation}
h(\vec r,t)=\sum_{a} \alpha_a \ h^{(a)}_{k}(r) \
{\cal K}^{(a)}_{kk_3}(\hat{r})\ e^{-i\varepsilon t}\ ,
\label{hfield}
\end{equation}
where the sum over $a$ accounts for
the possible ways of constructing the eigenstates of the same
grand spin and parity by combining the eigenstates of the spin,
isospin and orbital angular momentum, and the expansion
coefficients $\alpha_a$ are normalized by $\sum_a|\alpha_a|^2=1$.
Since we are assuming here that both the soliton and the heavy
mesons are infinitely heavy in the lowest energy state they should
be sitting one on top of the other at the same spatial point,
just propagating in time. That is, the radial functions $h^{(a)}_k(r)$
of the lowest energy eigenstate can be approximated by a
delta-function-like one, say $f(r)$, which is strongly peaked at
the origin and normalized as $\int dr\ r^2  |f(r)|^2 = 1$. Thus,
using orthonormalized eigenfunctions
${\cal K}^{(a)}_{kk_3}(\hat r)$ of the grand spin which satisfy
\begin{equation}
\int\!\!d\Omega {\,\mbox{Tr}}\left[
{\cal K}^{(a)}_{kk_3}(\hat{r})
 \bar{{\cal K}}^{(a')}_{k'k'_3}(\hat{r}) \right]
=-\delta_{aa'}\delta_{kk'}\delta_{k_3k'_3}\ ,
\label{Knorm}
\end{equation}
the field $h$ is normalized as
\begin{equation}
-\int\!\!d^3r {\,\mbox{Tr}}[h\bar{h}]=1\ .
\end{equation}

Replacing Eq.(\ref{hfield}) in Eq.(\ref{hveq}) and integrating out the radial part, we
obtain
\begin{equation}
\varepsilon\ {\cal K}_{kk_3}(\hat{r}) = \frac{g F^\prime(0)}{2}  \
{\cal K}_{kk_3}(\hat{r}) \left(2\vec\sigma\cdot\hat r\ \vec\tau\cdot\hat r-\vec\sigma\cdot\hat r\right)\ ,
\label{EoM1}
\end{equation}
with ${\cal K}_{kk_3} \equiv \sum_a \alpha_a \ {\cal K}^{(a)}_{kk_3}$. Here, we have used
that, near the origin, $F(r)\sim\pi+F^\prime(0)\ r$ and
consequently $\vec{A}\cdot\vec\sigma \sim \frac12 F^\prime(0)
(2\vec\sigma\cdot\hat r\vec\tau\cdot\hat r-\vec\sigma\cdot\hat r)$.

Thus, our problem is reduced to finding ${\cal K}_{kk_3}$.
For this purpose it is convenient to construct the grand spin
eigenstates ${\cal K}^{(a)}_{kk_3}(\hat{r})$ by combining the
eigenstates of the spin, isospin and orbital angular momentum.
Here, we construct first the eigenfunctions of
$\vec \Lambda = \vec L + \vec I$ by combining orbital
angular momentum and isospin eigenstates, and then
couple the resulting states to the spin eigenstates.
Since we are interested here in the lowest energy eigenmode of positive
parity, we can restrict the angular momentum $\ell$ to be 1. This statement
requires some explanation.
In general, when departures from a delta-like behavior are considered
the differential equations for the heavy meson radial functions
have a centrifugal term with a singularity $\ell_{eff}(\ell_{eff}+1)/r^2$
near the origin. Here, $\ell_{eff}$ is the ``effective" angular momentum\cite{CK85}
given by $\ell_{eff} = \ell\pm 1$ if $\lambda=\ell \pm 1/2$.
That behavior is due to the presence of a vector potential from the soliton
configuration $\vec{V}(\sim i(\hat{r}\times\vec{\tau})/r$, near the origin),
which alters the singular structure of $\vec{D}^2=(\vec{\nabla}-\vec{V})^2$
from $\ell(\ell+1)/r^2$ of the usual $\vec\nabla^2$ to $\ell_{eff}(\ell_{eff}+1)/r^2$.
Thus, the state with $\ell_{eff}=0$ can have most strongly peaked radial function
and become the lowest eigenstate. Note that $\ell_{eff}=0$ can be achieved only when
$\ell=1$. It is important to notice that combining the negative parity resulting
from this orbital wavefunction with the heavy meson intrinsic negative parity
we obtain that ground state heavy baryons have positive parity, as expected.
For $\ell=1$ two values of $\lambda$, $\frac12$ and $\frac32$, are possible.
Moreover, from the experience of the bound-state approach to strange hyperons,
where a similar situation arises\cite{CK85} , the lowest
energy state is expected to correspond to the lowest possible value
of $k$, i.e. $k=\frac12$.
Since we have $s =0, 1$ and $\lambda$=$\frac12$, $\frac32$, we can construct
three different grand spin states of $k=\frac12$.
Explicitly\cite{Min:1994qq},
\begin{eqnarray}
{\cal K}^{(1)}_{\frac12,\pm\frac12}({\hat r}) &=& \frac{1}{\sqrt{8\pi}}
\  \chi_\pm \ \vec\tau \cdot \hat r\ ,
\nonumber \\
{\cal K}^{(2)}_{\frac12,\pm\frac12}({\hat r}) &=& \frac{1}{\sqrt{24\pi}}
\  \chi_\pm  \ \vec\sigma \cdot \vec\tau \ \vec\tau \cdot \hat r \ ,
\\
{\cal K}^{(3)}_{\frac12,\pm\frac12}({\hat r}) &=& \frac{1}{\sqrt{48\pi}}
\  \chi_\pm  \ \left(\vec\sigma \cdot \vec\tau \
\vec\tau \cdot \hat r - 3\ \vec \sigma \cdot \hat r \right)\ .
\label{Ki}
\nonumber
\end{eqnarray}
Here, $ \chi_{+}=(0,-1)$ and $ \chi_{-}=(+1,0)$ are the isospin states
corresponding to $\bar{u}$ and $\bar{d}$, respectively.
The eigenstates ${\cal K}_{\frac12,\pm\frac12}({\hat r})$ of Eq.(\ref{EoM1}) can
be expanded in terms of these states
\begin{equation}
{\cal K}_{\frac12,\pm\frac12}({\hat r})
 = \sum_{a=1}^3 \alpha_a \ {\cal K}^{(a)}_{\frac12,\pm\frac12}({\hat r})\ ,
\label{ES}
\end{equation}
with the expansion coefficients given by the solution of the
secular equation
\begin{equation}
\sum_{b=1}^3{\cal M}_{ab}\ \alpha_b=-\ \varepsilon \ \alpha_a\ ,
\label{ESa}
\end{equation}
where the matrix elements ${\cal M}_{ab}$ are defined by
\begin{eqnarray}
{\cal M}_{ab} = \frac{gF^\prime(0)}{2} \int\!d\Omega {\,\mbox{Tr}}\left[ {\cal K}^{(a)}({\hat r})
\left(2\ \vec\sigma\cdot\hat r \ \vec\tau\cdot\hat r-\vec\sigma\cdot\hat r\right) \bar{{\cal K}}^{(b)}({\hat r})\ \right] \ .
\label{eme}
\end{eqnarray}
Note that the minus sign in Eq.~(\ref{ESa}) is due to the fact that the
basis states ${\cal K}^{(a)}_{\frac12,\pm\frac12}({\hat r})$ are
normalized as indicated in Eq.(\ref{Knorm}). With the explicit form of
${\cal K}^{(a)}_{\frac12,\pm\frac12}({\hat r})$ given by Eq.~(\ref{Ki}),
these matrix elements can be easily calculated.
The secular equation (\ref{ES})
yields three eigenstates. Since $g<0$ and
$F^\prime(0)<0$ (in the case of the
baryon-number-1 soliton solution), there is a
heavy-meson--soliton bound state of binding energy $-\frac32gF^\prime(0)$.
The two {\em unbound} eigenstates with positive eigenenergy
$+\frac12gF^\prime(0)$ are not consistent with the strongly peaked radial
functions. They are improper solutions of Eq.~(\ref{EoM1}).

In terms of the eigenmodes, the hamiltonian of the
system in the body fixed (i.e. soliton) frame has
the diagonal form
\begin{eqnarray}
H_{\rm bf} &=& M_{sol}
+ \sum_{n k k_3} \varepsilon_{n k} \ a_{n k k_3} \ a_{n k k_3}^\dagger = \nonumber \\
&=& M_{sol} + \varepsilon_{bs} \
\left( a_{+1/2}^\dagger\ a_{+1/2} + a_{-1/2}^\dagger\ a_{-1/2} \right) +...\ ,
\label{bsham}
\end{eqnarray}
where $n$ represents the extra quantum numbers
needed to completely specify a given eigenstate. Moreover, $a_{n k k_3}$  $(a_{n k k_3}^\dagger)$
are the usual meson annihilation (creation) operators. In the second line
of Eq.(\ref{bsham})
we have explicitly written the contribution of the bound state with
$\varepsilon_{gs} = -\frac32gF^\prime(0)$ found above, using the subscript $\pm1/2$ to indicate
the grand spin projection $k_3$.

What we have obtained so far is the heavy-meson--soliton bound state which
carries a baryon number and a heavy flavor.
Therefore, up to order $O(m_Q^0 N_c^{0})$ baryons containing a heavy quark such as
$\Lambda_Q$, $\Sigma_Q$ and $\Sigma_Q^*$ are degenerate in mass.
However, to extract physical heavy baryons of correct spin and isospin,
we have to go to the next order in $1/N_c$,
while remaining in the same order in $m_Q$, i.e. $O(m_Q^0 N_c^{-1})$.
This can be done by introducing time dependent $SU(2)$ collective variables
$C(t)$ associated with the degeneracy under simultaneous $SU(2)$ rotation
of the soliton configuration and the heavy meson fields
\begin{equation}
\xi(\vec{r},t) = C(t)\ \xi^{}_{\rm bf}(\vec{r})\ C^\dagger(t) \qquad \mbox{and} \qquad
h(\vec{r},t) = h_{\rm bf}(\vec{r},t) \ C^\dagger(t)\ ,
\label{CV}
\end{equation}
where $\xi^2_{\rm bf} \equiv U_0$, and then
performing the quantization by elevating them
to the corresponding quantum mechanical operators.
In Eq.~(\ref{CV}) and in what follows, $h_{\rm bf}$ refers to the heavy meson
field in the (isospin) soliton frame, while $h$ refers to that in the
laboratory frame, {\it i.e.}, the heavy quark rest frame.
Inserting Eq.~(\ref{CV}) in Eq.(\ref{leff})
we obtain an extra collective contribution of $O(m_Q^0 N_c^{-1})$ to the lagrangian
\begin{eqnarray}
L_{coll} = \frac12 \ {\cal I} \ \omega^2 + \vec Q \cdot \vec \omega \ ,
\label{Lcol}
\end{eqnarray}
where the ``angular velocity" $\vec\omega$  of the collective rotation is defined by
\begin{eqnarray}
C^\dagger\dot C \equiv  \frac{i}{2} \vec \tau \cdot \vec \omega\ ,
\label{Lcold}
\end{eqnarray}
${\cal I}$ is the moment of inertia of the rotating soliton, whose explicit
expression in terms of the soliton profile function $F(r)$
can be found in e.g. Refs.\cite{Sky62,Zahed:1986qz} ,
and
\begin{eqnarray}
\vec Q = - \frac{1}{4} \int d^3r
{\,\mbox{Tr}}\left[ h_{\rm bf} \left( \xi_{\rm bf}^\dagger \ \vec \tau \ \xi_{\rm bf} +
                     \xi_{\rm bf} \ \vec \tau \ \xi_{\rm bf}^\dagger \right) \bar{h}_{\rm bf}
   \right]\ .
\label{qcoll}
\end{eqnarray}

Taking the Legendre transform of the lagrangian
we obtain the collective hamiltonian as
\begin{eqnarray}
H_{coll} &=& \frac{1}{2{\cal I}}\left( \vec R- \vec Q \right)^2\ ,
\label{Hcol}
\end{eqnarray}
where $\vec R$ is the spin of the rotor given by $\vec R = {\cal I}\ \vec\omega + \vec Q$.

With the collective variable introduced as in Eq.~(\ref{CV}), the
isospin of the fields $U(x)$ and $h(x)$ is entirely shifted to $C(t)$.
To see this, consider the isospin rotation
\begin{equation}
U \rightarrow {\cal A}\ U \ {\cal A}^\dagger\ , \hskip 8mm
   h \rightarrow h\ {\cal A}^\dagger\ ,
\end{equation}
with ${\cal A}\in SU(2)_V$, under which the collective variables and fields
in the soliton frame transform as
\begin{equation}
C(t)\  \rightarrow {\cal A}\ C(t), \hskip 8mm
h_{\rm bf}(x) \ \rightarrow \ h_{\rm bf}(x)\ .
\end{equation}
Thus, the $h$-field is isospin-blind in the (isospin) soliton frame.
The conventional Noether construction gives the isospin of the system,
\begin{equation}
I^a = \frac12 {\,\mbox{Tr}} \left[ \tau^a C \tau^b C^\dagger \right] \,
\left( {\cal I}\ \omega^b + Q^b \right) = D^{ab}(C) R^b\ ,
\end{equation}
where $D^{ab}(C)$ is the adjoint representation of the $SU(2)$ transformation
associated with the collective variables $C(t)$.

The eigenfunctions of the rotor-spin operator  are the usual
Wigner ${\cal D}$-functions. In terms of these eigenfunctions and the heavy meson
bound states $|\pm 1/2 \rangle_{bs}$, the heavy baryon state of isospin $i_3$ and spin
$s_3$ containing a heavy quark can be constructed as
\begin{equation}
|i; i_3,s_3\rangle = \sqrt{2 i + 1} \sum_{m=\pm 1/2}
(i,s_3-m,1/2,m|1/2,s_3)\  {\cal D}^{(i)}_{i_3,-s_3+m}(C)\ |m\rangle_{bs}\ ,
\label{HBS}
\end{equation}
where $i=0$ for $\Lambda_Q$ and $i=1$ for $\Sigma_Q$ and $\Sigma^*_Q$.

Treating the collective Hamiltonian (\ref{Hcol}) to first order in
perturbation theory we obtain
\begin{equation}
m_i = M_{sol} + \varepsilon_{bs}^{} + \frac{1}{2{\cal I}} \Big( i (i + 1) + 3/4 \Big)
\ .
\end{equation}
Here, we have used that explicit evaluation shows
\cite{Oh:1994yv}
\begin{eqnarray}
{_{bs}\langle m | \vec Q |m\rangle_{bs}} &=& 0\ ,
\label{cfac} \\
{_{bs}\langle m |\vec Q^2 |m\rangle_{bs}} &=& 3/4 \ .
\label{qfac}
\end{eqnarray}
These two results deserve some comments. First we note that general use
of the Wigner-Eckart theorem implies
\begin{equation}
< n,k,k_3  | \vec Q | n,k,k'_3 > = c_{n k}
< n,k,k_3  | \vec K | n,k,k'_3 > \ .
\label{WETh}
\end{equation}
The constants $c_{n k}$ are usually called "hyperfine splitting" constants.
Eq.(\ref{cfac}) implies that for the ground state $c_{gs}=0$ in the heavy
quark limit. As a consequence of this, the Hamiltonian depends only on the rotor-spin
so that $\Sigma_Q$ and $\Sigma^*_Q$ become degenerate as expected from the heavy quark
symmetry. It is clear that corrections that imply departures from heavy quark limit
could lead to non-vanishing values of $c_{gs}$. It is also important to notice
that to obtain the result Eq.(\ref{qfac}) one should take into account
all possible intermediate states.

In order to compare the results with experimental
heavy baryon masses, we have to add the heavy meson masses subtracted so far
from the  eigenenergies. The mass formulas to be compared with data are
\begin{eqnarray}
m^{}_{\Lambda_Q^{}}& = &M_{sol} + \overline{m}_\Phi^{} - \frac32 gF^\prime(0) + \frac{3}{8{\cal I}}\ ,
\nonumber \\
m^{}_{\Sigma_Q^{}} = m^{}_{\Sigma^*_Q}
 &=& M_{sol} + \overline{m}_\Phi^{} - \frac32 gF^\prime(0) + \frac{11}{8{\cal I}}\ ,
\label{HBmass1}
\end{eqnarray}
where $\overline{m}_\Phi^{}$ is the weighted average mass of the heavy meson multiplets,
$\overline{m}_\Phi^{}~=~(3m^{}_{\Phi^*}+m^{}_\Phi)/4$.
In the case of $Q=c$, we have $\overline{m}_\Phi^{}=1973$~MeV while for
$Q=b$, $\overline{m}_\Phi^{}=5314$~MeV .
The $SU(2)$ quantities $M_{sol}$ and
${\cal I}$ are obtained from the nucleon and $\Delta$
masses
\begin{equation}
M_{sol}=866\mbox{~MeV}, \qquad \mbox{and} \qquad 1/{\cal I}=195\mbox{~MeV}\ .
\end{equation}
Finally, the unknown value of $gF^\prime(0)$ can be adjusted to fit the
observed value of the $\Lambda_c$ mass,
\begin{equation}
m^{}_{\Lambda^{}_c}=2286\mbox{ MeV}=M_{sol}+\overline{m}_\Phi^{}-\frac32 gF^\prime(0) + \frac{3}{8{\cal I}}\ ,
\end{equation}
which implies that
\begin{equation}
gF^\prime(0) = 417 \mbox{ MeV}\ .
\end{equation}
This leads to a prediction on the
$\Lambda_b$ mass and the average masses of the
$\Sigma^{}_Q$-$\Sigma^*_Q$ multiplets,
$\overline{m}^{}_{\Sigma^{}_Q} [\equiv
\frac13(2m^{}_{\Sigma^*_Q}+m^{}_{\Sigma^{}_Q})]$,
\begin{eqnarray}
m_{\Lambda^{}_b} &=& M_{sol} + \overline{m}_B - \frac32gF^\prime(0)
+3/8{\cal I} = 5627 \mbox{ MeV} \ , \\
\overline{m}^{}_{\Sigma^{}_c} &=& M_{sol} + \overline{m}_D
- \frac32gF^\prime(0) + 11/8{\cal I} = 2481 \mbox{ MeV} \ ,
\\
\overline{m}^{}_{\Sigma^{}_b} &=& M_{sol} + \overline{m}_B
- \frac32gF^\prime(0) + 11/8{\cal I} = 5822 \mbox{ MeV}\ .
\end{eqnarray}
These are comparable with the experimental masses
\cite{Amsler:2008zzb} of $\Lambda_b$ (5620~MeV),
$\Sigma_c$~(2454~MeV), $\Sigma^*_c$ (2518~MeV),
$\Sigma_b$ (5811~MeV) and $\Sigma^*_b$ (5833~MeV).
Furthermore, with the Skyrme lagrangian (with the quartic term for
stabilization), the wavefunction has a slope
$F^\prime(0)\sim -2ef_\pi\approx -700$ MeV
near the origin, which implies $g\sim -0.6$.
This is also consistent with the values given at the end of
the previous subsection.

The role of light vector mesons in the description of the heavy-meson--soliton
system was analyzed in Ref.\cite{Gupta:1993kd}$\!$ . In fact, using effective heavy
quark symmetric lagrangians that incorporate light vector mesons
\cite{Casalbuoni:1992gi,Schechter:1992ue}, it was shown that the effect of these
light degrees of freedom could be relevant. Within this scheme the extension of
the light flavor group to SU(3) was also considered\cite{Momen:1993ax}.

Up to now, we have discussed how one can obtain the heavy baryon states
containing a heavy quark, $\Sigma_Q^{}$, $\Sigma^*_Q$ and $\Lambda_Q^{}$,
as heavy-meson--soliton bound states treated in the standard way:
a heavy-meson--soliton bound state is first found and then quantized by rotating
the whole system in the collective coordinate quantization scheme. This amounts
to proceeding systematically in a decreasing order in $N_c$; i.e.: in the first step
only terms up to $N_c^0$ order are considered, in the next step terms of order
$1/N_c$ are also taken into account, etc.
In this way of proceeding, the heavy mesons first lose their quantum numbers
(such as the spin and isospin), with only the grand spin preserved.
The good quantum numbers are recovered when the
whole system is quantized properly.
An alternative approach was adopted in Ref.~\cite{Jenkins:1992zx}.
In this approach,
the soliton is first quantized to produce the light baryon states such as nucleons
and $\Delta$'s with correct quantum numbers. Then, the heavy mesons with explicit
spin and isospin are coupled to the light baryons to form heavy baryons as a bound
state. Compared with the traditional one which is a ``soliton body-fixed" approach,
this approach may be interpreted as a  ``laboratory-frame" approach.
It has been shown\cite{Min:1994qq} , however, that both approaches lead to the same
results in the heavy quark limit.

It should be stressed that in the heavy quark limit discussed so far
one cannot account for the experimentally observed hyperfine splittings,
like e.g. the $\Sigma^*_c$-$\Sigma_c$ mass difference. Another consequence of taking
such limit is the existence of parity doublets in the spectrum
of the low-lying excited states\cite{Oh:1994yv,Schechter:1994ip}.
This follows from the fact that in the heavy quark limit the centrifugal
barrier that would affect states with $\ell_{eff} > 0$ plays no role.
It is clear that finite heavy quark mass corrections have to be taken into
account in order to have a more realistic description of the heavy baryon
properties in the present topological soliton framework. How to account
for such corrections will be discussed in the following section.

\section{Beyond the Heavy Quark Limit}
\label{beyhql}

In the previous section, we have limited ourselves to the heavy quark limit.
Thus, heavy baryon masses have been computed to leading order in $1/m_Q$,
that is to $O(m_Q^0)$. Here, we will consider the corrections
implied by the use of finite heavy quark masses.

The $\Sigma^*_Q$-$\Sigma_Q$ mass difference due to the leading heavy quark
symmetry breaking was first computed in Ref.\cite{Jenkins:1992ic} using the
alternative method mentioned at the end of Subsec.\ref{ubsa}.
As an illustration of the equivalence of the two approaches, we briefly
discuss how the corresponding results can be obtained using the soliton body fixed
approach described at length in that subsection. The leading order lagrangian in
the derivative expansion that breaks the heavy quark symmetry is\cite{Wis92}
\begin{equation}
{\cal L}_1 = \frac{\lambda_2}{m_Q}
{\,\mbox{Tr}} \left[ \sigma^{\mu\nu} H \sigma_{\mu\nu} \bar{H} \right]\ ,
\label{L1}
\end{equation}
which leads to a $\Phi^*$-$\Phi$ mass difference
\begin{equation}
m_{\Phi^*} - m_\Phi = - \frac{8\lambda_2}{m_Q}\ .
\label{depi}
\end{equation}
Assuming as in Subsec.\ref{ubsa} that the radial functions are peaked
strongly at the origin, the inclusion of this heavy quark symmetry breaking
lagrangian implies that the equation of motion Eq.(\ref{hveq}) gets an
additional term. Namely, one obtains
\begin{equation}
i\ \dot h = g\ h\  \vec A \cdot \vec\sigma
   + \frac{2\lambda_2}{m_Q}\ \vec\sigma \cdot \left( h\ \vec\sigma\right) \ .
\end{equation}
One can now consider the last term as a perturbation and compute its effect
on the $k=1/2$ bound state.
Since ${\cal L}_1$ breaks only the heavy quark spin symmetry the grand
spin is still a good symmetry of the equation of motion. Thus, the
eigenstates can be classified by the corresponding quantum numbers.
Expanding in terms of the three possible basis
states ${\cal K}^{(a)}_{\frac12 k_3}$ given in Eq.(\ref{Ki})
the problem reduces to finding the solution of the secular equation
\begin{equation}
\sum_{b=1}^3\left( {\cal M}_{ab} + \delta {\cal M}_{ab} \right) \alpha_b = - \varepsilon \ \alpha_a\ ,
\label{M0M1}
\end{equation}
with ${\cal M}_{ab}$ given by Eq.(\ref{eme}) and
\begin{equation}
\delta {\cal M}_{ab} = \frac{2\lambda_2}{m_Q} \int d\Omega {\,\mbox{Tr}} \left[
\vec\sigma \cdot \left( {\cal K}^{(a)}_{\frac12 k_3} \vec\sigma \right) \bar{{\cal K}}^{(b)}_{\frac12 k_3} \right] \ .
\label{M0M2}
\end{equation}
It turns out that up to first order in perturbation, the bound state energy remains
unchanged while the corresponding eigenstate ${\cal K}^{bs}_{\frac12 k_3}$ is perturbed to
\begin{equation}
{\cal K}^{bs}_{\frac12 k_3} =
\frac{1}{2} \ (1 + 3\ \kappa ) \  {\cal K}^{(1)}_{\frac12 k_3}
- \frac{\sqrt3}{2} \ (1 - \kappa) \ {\cal K}^{(2)}_{\frac12 k_3}\ ,
\label{BSptbd}
\end{equation}
with
\begin{equation}
\kappa = - \frac{\lambda_2}{m_Q} \frac{1}{gF^\prime(0)}\ .
\end{equation}

The heavy baryons can be obtained by quantizing the heavy-meson--soliton
bound state in the same way as explained in Subsec.\ref{ubsa}. It leads
to the heavy baryon states of Eq.~(\ref{HBS}) with $|m\rangle_{bs}$ replaced
by the perturbed state of Eq.~(\ref{BSptbd}).
Due to the perturbation, the expectation value of $\vec Q$
defined by Eq.~(\ref{qcoll}) with respect to the bound states
does not vanish. In fact, one gets that the hyperfine constant
is given by
\begin{equation}
c = 2\epsilon = - \frac{2\lambda_2}{m_Q} \frac{1}{gF^\prime(0)}\ .
\label{cp}
\end{equation}
With the help of Eq.~(\ref{WETh}), one can compute the expectation value of
the collective hamiltonian (\ref{Hcol})
\begin{equation}
m_{i,j} = M_{sol} + \varepsilon_{bs} + \frac{1}{2{\cal I}}
\Big( (1-c)i(i+1) + cj(j+1) - ck(k+1) + \frac34 \Big)\ .
\label{massform}
\end{equation}
Thus, the $\Sigma_Q^*$-$\Sigma_Q$ mass difference
is obtained as
\begin{equation}
m_{\Sigma^*_Q} - m_{\Sigma_Q} =  \frac{3 c}{2{\cal I}} =
\frac{(m_\Delta - m_N)(m_{\Phi^*} - m_\Phi)}{4gF^\prime(0)}\ ,
\end{equation}
where Eqs.(\ref{depi}) and (\ref{cp}) together with the
resulting expression for the $\Delta$-$N$ mass splitting
in terms of the moment of inertia $\cal I$ have been used.
Note that the mass splittings have the dependence on
$m_Q$ and $N_c$ that agrees with the constituent quark model.
The $\Phi^*$-$\Phi$ mass difference is of order $1/m_Q$ and
the $\Delta$-$N$ mass difference is of order $1/N_c$.
This implies that the $\Sigma^*_Q$-$\Sigma_Q$ mass difference is of
order $1/(m_QN_c)$. Substituting $gF^\prime(0) = 417$ MeV, we obtain
\begin{equation}
m_{\Sigma_c^*} - m_{\Sigma_c} = 25 \mbox{ MeV}  \qquad \mbox{and} \qquad
m_{\Sigma_b^*} - m_{\Sigma_b} = 8 \mbox { MeV}\ .
\end{equation}
The experimentally measured $\Sigma_c^*$-$\Sigma_c$
mass difference $\sim 64$ MeV is about three times larger than this
Skyrme model prediction. Something similar happens in
the case of the $\Sigma_b^*$-$\Sigma_b$
mass difference, the empirical value of which is $\sim 21$ MeV.

This failure to reproduce the observed hyperfine splittings
naturally suggests the need for including additional heavy-spin violating terms,
of higher order in derivatives. However, since there are many possible
terms with unknown coefficients such a systematic perturbative approach
turns out not to be very predictive. To overcome this problem some relativistic
lagrangian models written in terms of the ordinary pseudoscalar and vector
fields (rather than the heavy fluctuation field multiplet Eq.(\ref{H}))
have been used. A typical model of this type which only includes
pseudoscalar fields in the light sector is given by
\begin{eqnarray}
\!\!\!\!\!\!\!\!\!\!\!\!\!\!\!\!\!{\cal L} &=& {\cal L}_l^{Sk}
+ D_\mu \Phi (D^\mu \Phi)^\dagger - m^2_\Phi \Phi \Phi^\dagger
- \frac12 \Phi^{*\mu\nu} \Phi^{*\dagger}_{\mu\nu} + m^2_{\Phi^*} \Phi^{*\mu} \Phi^{*\dagger}_\mu
\nonumber \\
& &\!\!\! + f_{_Q} ( \Phi A^\mu \Phi^{*\dagger}_\mu + \Phi^*_\mu A^\mu \Phi^\dagger) +
\frac{g_{_Q}}{2} \ \varepsilon^{\mu\nu\lambda\rho} (\Phi^*_{\mu\nu} A_\lambda \Phi^{*\dagger}_\rho
+ \Phi^*_\rho A_\lambda \Phi^{*\dagger}_{\mu\nu})\ ,
\label{lagrangian}
\end{eqnarray}
where $D_\mu \Phi = \partial_\mu \Phi - \Phi V_\mu^\dagger$,
$\varepsilon_{0123} = +1$, and
$f_{_Q}$ and $g_{_Q}$ are the $\Phi^* \Phi M$ and $\Phi^* \Phi^* M$ coupling
constants, respectively. The field strength tensor is
defined in terms of the covariant derivative
$D_\mu \Phi^*_\nu = \partial_\mu \Phi^*_\nu - \Phi^*_\nu V_\mu^\dagger$
as
\begin{equation}
\Phi^*_{\mu\nu} = D_\mu \Phi^*_\nu - D_\nu \Phi^*_\mu \ ,
\end{equation}
and the vector $V_\mu$ and axial vector $A_\mu$ have been defined in
Eq.(\ref{currents}). In principle, Eq.(\ref{lagrangian}) has two
independent coupling constants $f_{_Q}$ and $g_{_Q}$. However, in order
to respect heavy quark symmetry they should be related to
each other as
\cite{Yan92}
\begin{equation}
\lim_{m_Q\rightarrow\infty}f_{_Q} / 2 m_{\Phi^*}
= \lim_{m_Q\rightarrow\infty}g_{_Q} = g\ ,
\label{hqr}\end{equation}
where $g$ is the universal coupling constant appearing in Eq.(\ref{LvHx}).
It should be noted that even to order $1/m_Q$,  Eq.(\ref{lagrangian})
leads to extra contributions to the hyperfine splittings\cite{Harada:1996wv}.

The interacting heavy-meson--soliton system described by the lagrangian
Eq.(\ref{lagrangian}) can be treated following a procedure similar to
the one described at length in Subsec.\ref{ubsa}. It should be noted,
however, that the need to treat the finite mass corrections non-perturbatively
implies that departures from a $\delta$-like behaviour of the heavy meson
radial wavefunctions should be taken into account. Thus, the equations of
motions which describe the dynamics of the heavy mesons moving in
the static soliton background field should be solved numerically.
It turns out that, for a given value of $g$, the binding energies are somewhat
smaller than the ones obtained in the heavy quark limit\cite{Oh:1994vd}. Concerning
the hyperfine
splittings, although the use of the effective lagrangian Eq.(\ref{lagrangian})
leads to some improvement, it is not still
sufficient to bring the predicted $\Sigma^*_Q$-$\Sigma_Q$ mass splitting
into agreement with experiment. The prediction for such a splitting
is actually correlated to those for the $\Sigma_Q$ - $\Lambda_Q$ and
$\Delta$-$N$ splittings according to
\begin{equation}
m_{\Sigma^*_Q} - m_{\Sigma_Q} =
m_\Delta - m_N - \frac32 (m_{\Sigma_Q} - m_{\Lambda_Q})\ .
\label{desig}
\end{equation}
This formula follows from Eq.(\ref{massform}), and depends only
on the collective quantization procedure being used rather than
on the detailed structure of the model. If $m_\Delta - m_N$ and
$m_{\Sigma_c} - m_{\Lambda_c}$ are taken to agree with their
empirical value, Eq.(\ref{desig}) predicts $42$ MeV rather than
the empirical value $64$ MeV. In the case of the bottom baryons
one gets $6$ MeV to be compared to the empirical value $ 21$ MeV.
This means that, within the present quantization framework, it is
not possible to exactly predict
all the three mass differences appearing in Eq.(\ref{desig}).
Thus, the goodness of the approach must be judged by
looking at the overall predictions for the heavy baryon masses.

\begin{figure}[bp]
\centerline{\epsfig{file=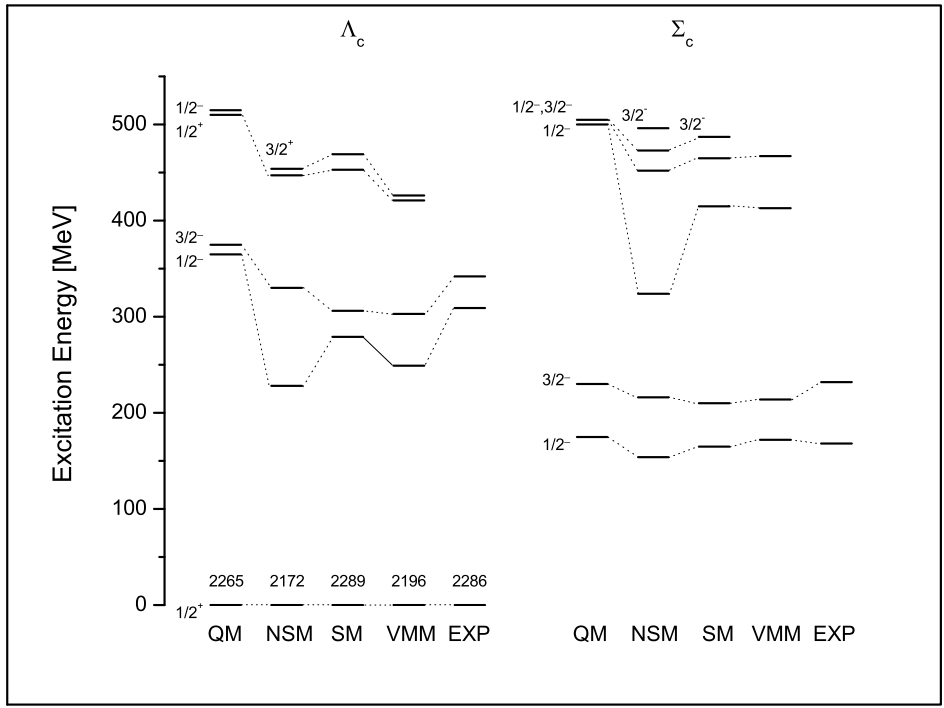,width=14cm,angle=0}}
\caption{Excitation spectra of charm baryons in soliton models as compared to the results of the
quark model (QM) of Ref.\cite{Capstick:1986bm} and the present empirical data\cite{Amsler:2008zzb} (EXP).
NSM corresponds to the soliton model calculation of Ref.\cite{RS91} where heavy quark symmetry
has not been explicitly implemented. SM and VMM refer to soliton models
which incorporate heavy quark symmetry. SM corresponds to a calculation\cite{Oh:1995ey}
where only pseudoscalars have taken into account in the light sector, while VMM to the
calculation of Ref.\cite{Schechter:1995vr} where light vector mesons have been also explicitly
included. The numbers above the lowest $\Lambda_c$ state correspond to the absolute masses (in MeV)
of this state.}
\label{fig1}
\end{figure}

\begin{figure}[bp]
\centerline{\epsfig{file=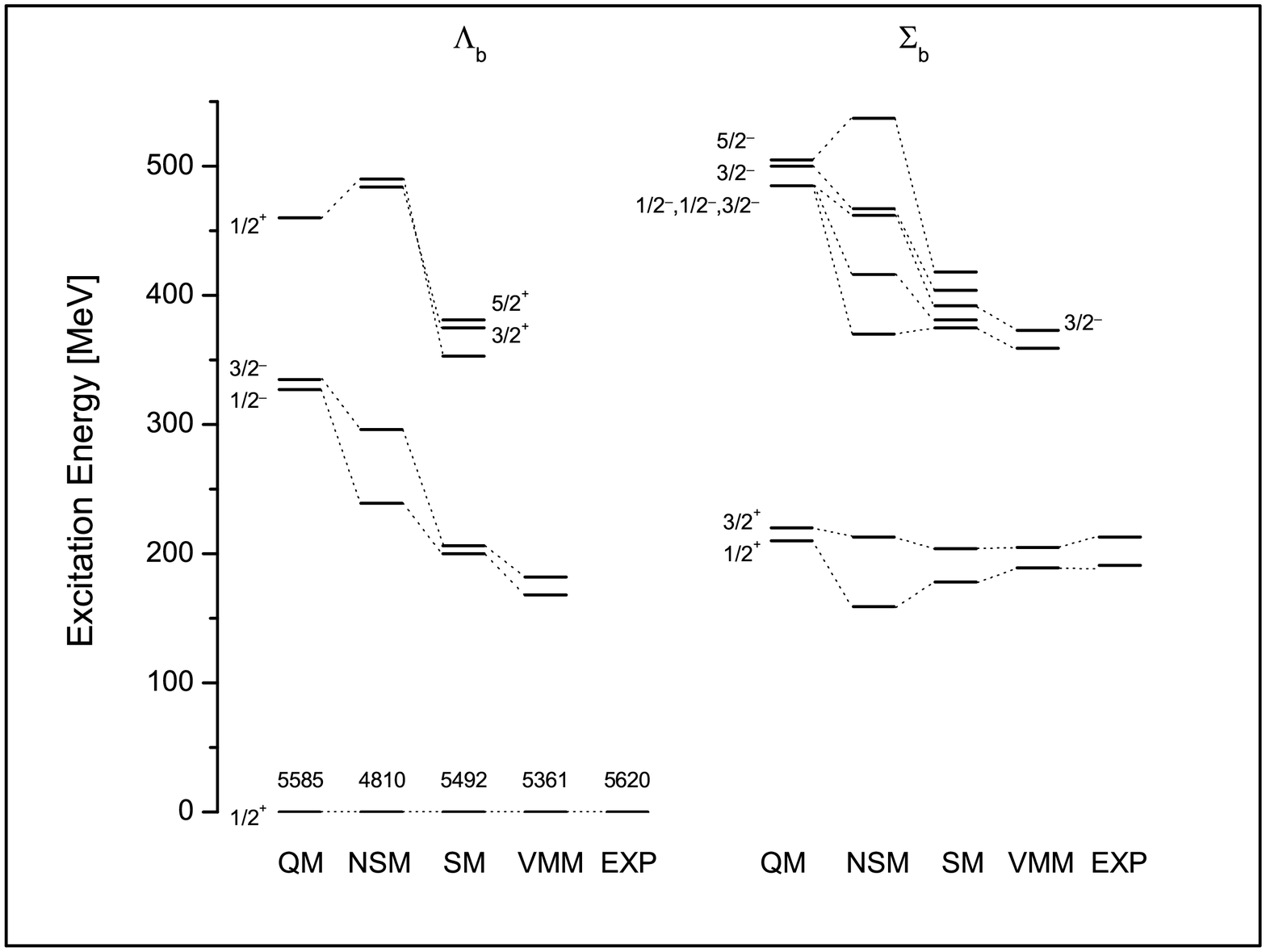,width=14cm,angle=0}}
\caption{Excitation spectra of bottom baryons. Notation as in Fig.\ref{fig1}.}
\label{fig2}
\end{figure}

In this context, the study of possible excited states turns out to be
of great interest. As already mentioned, in the heavy quark limit degenerate
doublets of excited states are obtained. However, such limit implies
that both the soliton and the heavy mesons are infinitely heavy sitting
one on top of the other. It is evident that, due to the ignorance
of any kinetic effects, this approximation is not expected to
work well for the orbitally and/or radially excited states.
In Ref. \cite{Chow:1994sz} the kinetic effects due to the finite
heavy meson masses were estimated by approximating their static potentials
by a quadratic form with the curvature determined at the origin.
Such a harmonic oscillator approximation is valid only when the heavy
mesons are sufficiently massive so that their motions are restricted to
a very small range. The situation was somewhat improved in Ref. \cite{Schechter:1994ip}
by solving an approximate Schr\"odinger-like equation and incorporating the
light vector mesons. In the context of the model defined by Eq.(\ref{lagrangian}),
in which only pseudoscalar degrees of freedom are present in the light sector,
the exact solution of the equations of motion of the heavy meson bound states
were first obtained in Ref.\cite{Oh:1994ux} and their collective coordinate quantization
performed in Ref.\cite{Oh:1995ey}. The typical resulting excitation spectra for
the low-lying charm and bottom baryons obtained from these calculations
(SM) are shown in Figs.\ref{fig1} and \ref{fig2}, respectively.

For comparison, we
also include in these figures the results of the quark model (QM) calculation of Ref.\cite{Capstick:1986bm}
(more recent quark model calculations \cite{Migura:2006ep} lead to qualitatively similar results),
those resulting from naive extension\cite{RS91} of the bound state approach to the strangeness
(NSM) and the empirically known values\cite{Amsler:2008zzb} (EXP).
Note that the excitation
energies are taken with respect to the mass (also indicated in the figures) of the
lowest $\Lambda_c$ and $\Lambda_b$, respectively. Finally, in order to see
the impact of including the light vector mesons in the effective lagrangian,
the excitation spectra resulting from the calculations of Ref.\cite{Schechter:1995vr}
(VMM) are also displayed.

In the case of the charm sector, we observe that the predictions
for the absolute values of the ground state $\Lambda_c$ mass are similar in all
soliton models calculations, and are in reasonable agreement with its empirical value
and the QM prediction.
As for the low lying spectra, we see that they are all qualitatively similar. From a more
quantitative point of view, the SM version of the skyrmion models seems to provide a more accurate
description of the splitting between the two lowest lying negative parity excited $\Lambda_c$
baryons, although the corresponding centroid is somewhat underestimated as compared with
present experimental results. In any case, for these particular states the soliton models
based on heavy quark symmetry certainly do better than the QM of Ref.\cite{Capstick:1986bm}
and the soliton calculation NSM. For the $\Sigma_c$ baryons, the predictions
of the SM and VMM results are very similar with the main difference, with respect to the QM
and NSM predictions, being the position of the second $1/2^-$ state.
Concerning the bottom sector, looking at the absolute value of the ground state $\Lambda_b$, we clearly
see that the NSM tends to grossly overestimate the bottom meson binding energy. In this sense,
although as discussed below the inclusion of other effects might still be required, the soliton models
based in heavy quark symmetry (SM and VMM) lead to predictions which are in much better agreement with
the empirical values. As for the excitation spectra, we see that all the models predict a
similar ordering of low-lying states. However, the only two excitation energies that can be compared
with existing empirical data, i.e. those corresponding to the $\Sigma_b$ and $\Sigma_b^*$, are also
much better reproduced by the SM and VMM results. It should be noticed that those models also predict
rather small excitation energies ($\approx 200$ MeV) for the lowest lying negative $1/2^-$ and
$3/2^-$ states as compared with the QM prediction (above $300$ MeV).

Another kinetic correction that has to be taken into account
is related to the recoil effects due to the finite soliton
mass. This type of effect has been considered in several
works\cite{Schechter:1994ip,Oh:1994ux,Schechter:1995vr,Oh:1997tp,Cohen:2007up}.
As expected, they tend to decrease the heavy-meson--soliton binding
energies leading to predictions which, particularly in the case
of bottom baryons, are in better agreement with empirical data.

It should be mentioned that in the combined heavy quark and large $N_c$ limit
a dynamical symmetry connecting excited heavy baryon states with the corresponding
ground states exists\cite{Chow:1999hm}. Assuming that such symmetry holds
as an approximate symmetry at finite values of $m_Q$ and $N_c$ one can
develop an effective theory formulated in terms of the expansion parameter
$\lambda \sim 1/m_Q, 1/N_c$. Within such scheme, up to next-to-leading order
an average excitation energy of $\sim 300$ MeV is obtained for the first
negative parity $\Lambda_b$ excited states. Such value is somewhat larger
than the one obtained within heavy-meson--soliton bound state models,
as it can be seen from Fig.\ref{fig2}.

We conclude this section by mentioning that, in addition to the masses,
other heavy baryon properties have been studied
using the heavy-meson--soliton bound state picture. For example, magnetic
moments have been analyzed in the heavy quark limit\cite{Oh:1995eu}
and beyond it\cite{Scholl:2003ip}. The radiative decays of excited
$\Lambda_Q$ have been also considered\cite{Chow:1995nw}. Finally,
the possible existence of multibaryons with heavy flavors\cite{Schat:1999dw,Kopeliovich:1999cu}
and other exotic states\cite{Riska:1992qd,Oh:1994ky,Bander:1994sp} have also been
investigated.

\section{Relation with the bound state approach to strangeness}
\label{relbsa}

Thus far, we have discussed in detail a description of heavy
baryons in which one begins from the heavy quark symmetry limit
and then consider deviations from such a limit which start
with order $1/m_Q$ corrections. However, as mentioned in the
introduction, the picture proposed in Ref.\cite{RS91}
in which the heavy quark regime is approached from below, i.e.
starting form a chiral invariant lagrangian and accounting
for the heavy meson mass effects by the inclusion of suitable
symmetry breaking terms, also turns out to be, at least qualitatively,
successful. Therefore, it is interesting to see whether it is
possible to find a dynamical scheme which allows to go continuously from
the chiral regime to the heavy quark regime.

Suppose that one starts with three massless quarks, assuming the spontaneous
breaking of chiral $SU(3)_L\times SU(3)_R$ down to the $SU(3)_V$ vector
symmetry. The chiral field can be written as
\begin{equation}
U=\exp\left[\frac{i}{f_\pi} \left(  \begin{array}{ccc}
\pi^0+\frac1{\sqrt3}\Psi & \sqrt2\pi^+       & \Phi^+       \\
\sqrt2\pi^-              &  - \pi^0 + \frac1{\sqrt3}\Psi  & \Phi^0 \\
\Phi^-                   &  \bar{\Phi}^0     & -\frac{2}{\sqrt3}\Psi
\end{array}\right)\right]\ .
\label{U3}
\end{equation}
Here, $\Phi^+$, $\Phi^0$, $\Phi^-$, $\bar{\Phi}^0$ and $\Psi$ denote the mesons
with the quantum numbers of $\bar{h}\gamma_5 u$, $\bar{h}\gamma_5 d$,
$\bar{u}\gamma_5 h$ and $\bar{d}\gamma_5 h$ and
$\bar{u}\gamma_5u+\bar{d}\gamma_5d-2\bar{h}\gamma_5h$, respectively.
For example, if $h$=$s$, they correspond to $K^+$, $K^0$, $K^-$,
$\bar{K}^0$ and $\eta_8$. The effective action can be obtained
by adding the Wess-Zumino term\cite{Wess:1971yu} $\Gamma_{WZ}$
to the lagrangian for interactions among the Goldstone bosons given by
generalizing Eq.(\ref{Lsk}) to three flavors. Namely,
\begin{equation}
\Gamma  = \int d^4x \ {\cal L}^{Sk}_l + \Gamma_{WZ}\ .
\label{Lsu3}
\end{equation}
The Wess-Zumino term cannot be written as a local lagrangian
density in $(3+1)$ dimensions. However, it can be expressed
as a local action in five-dimensions\cite{Witten:1983tw},
\begin{equation}
\Gamma_{WZ}=-\frac{iN_c}{240\pi^2}\int_{M_5}\!\!d^5x\
\varepsilon^{\mu\nu\rho\sigma\lambda}{\,\mbox{Tr}}\left[ U^\dagger\partial_\mu U
U^\dagger\partial_\nu U U^\dagger \partial_\rho U U^\dagger
\partial_\sigma U U^\dagger\partial_\lambda U\right],
\label{LWZ}
\end{equation}
where the integration is over a five-dimensional disk whose boundary is
the ordinary space-time $M_4$ and $U$ is extended so that $U(\vec{r},t,s=0)
=1$ and $U(\vec{r},t,s=1)=U(\vec{r},t)$.  This term is non-vanishing for
$N_f\geq 3$. When the soliton is built in $SU(2)$ space, this
term does not contribute. However, we shall be considering (2+1) flavors
where one flavor can be heavy, in which case the dynamics
can be influenced by the Wess-Zumino term as in the Callan-Klebanov (CK)
model\cite{CK85}.
What we are interested in is the situation where the symmetry
$SU(3)_L\!\times\!SU(3)_R$ is {\it explicitly} broken to
$SU(2)_L\!\times\!SU(2)_R\!\times\!U(1)$ by an $h$-quark
mass,
thereby making the $\Phi$-meson massive and its decay constant $f_\Phi$
different from that of the pion. These two symmetry
breaking effects can be effectively incorporated into the
lagrangian by a term of the form\cite{RS91}
\begin{eqnarray}
{\cal L}_{_{sb}}&=&\frac16 f_\Phi^2
m_\Phi^2{\,\mbox{Tr}}[(1-\sqrt3\lambda_8)(U+U^\dagger-2)]
\nonumber \\
& & + \frac1{12}(f_\Phi^2-f_\pi^2){\,\mbox{Tr}}[(1-\sqrt3\lambda_8)
(U\partial_\mu U^\dagger\partial^\mu U
+U^\dagger\partial_\mu U\partial^\mu U^\dagger)]\ ,
\label{Lsb}
\end{eqnarray}
where, for simplicity, we turn off the light quark masses. The appropriate ansatz
for the chiral field is the CK-type which we shall take in the form
\begin{eqnarray}
U= N_\pi \ N_\Phi \ N_\pi\ ,
\label{Uck}
\end{eqnarray}
where $N_\pi = \mbox{diag}\left(\xi,1\right)$, with the $SU(2)$ matrix $\xi$ defined by
Eq.~(\ref{xi}),  and
\begin{equation}
N_\Phi =
\exp\left[ \frac{i\sqrt2}{f_\pi}
\left(
\begin{array}{cc}
{\bf 0}  & \Phi^\dagger \\
\Phi        &  0
\end{array}
\right) \right] \ ,
\end{equation}
with the $\Phi$-meson
anti-doublets $\Phi=(\Phi^-\!\!,\bar{\Phi}^0)$ and doublets
$\Phi^\dagger=(\Phi^+\!\!, \Phi^0)^T$.

Substituting the CK ansatz (\ref{Uck}) into the action (\ref{Lsu3})
with the symmetry breaking term (\ref{Lsb}) and expanding up to
second order in the $\Phi$-meson field, we obtain
\begin{equation}
{\cal L}={\cal L}^{Sk}_l + D_\mu \Phi (D_\mu \Phi)^\dagger - M_\Phi^2 \Phi \Phi^\dagger
-\Phi A_\mu^\dagger A^\mu \Phi^\dagger
-\frac{iN_c}{4f_P^2}B_\mu \Big(D^\mu \Phi \Phi^\dagger-\Phi (D^\mu \Phi)^\dagger\Big),
\label{Lck}
\end{equation}
where we have rescaled the $\Phi$-meson fields as $\Phi/\kappa$ with $\kappa =
f_\Phi / f_\pi$.
The covariant derivative
$(D_\mu \Phi)^\dagger$ is $(\partial_\mu+V_\mu)\Phi^\dagger$, the
vector field $V_\mu$ and the axial-vector field $A_\mu$ are
the same as in the lagrangian (\ref{lagrangian}), and $B_\mu$
is the topological current
\begin{equation}
B^\mu=\frac{1}{24\pi^2}\varepsilon^{\mu\nu\lambda\rho}{\,\mbox{Tr}}\left[U^\dagger\partial_\nu U
U^\dagger\partial_\lambda UU^\dagger\partial_\rho U\right],
\end{equation}
which is the baryon number current in the Skyrme model.

With the identification $\Phi = K$, the lagrangian Eq.(\ref{Lck}) has been
successfully used in the strange sector. In fact, using the empirical
values for $m_K$ and the $f_K/f_\pi$ ratio this lagrangian leads to a kaon-soliton bound
state which allows for a very good description of the strange hyperon
spectrum\cite{RS91}, once an $SU(2)$ collective quantization similar to
the one described in Subsec.\ref{ubsa} is performed. Moreover,
the existence of an excited $\ell=0$ state provides a natural
explanation for the negative parity $\Lambda(1405)$
hyperon\cite{CK85,Schat:1994gm}. The results displayed at the
end of Sec.\ref{beyhql} (those labelled NSM in Figs.1 and 2) show that
the straightforward extension of this approach\cite{RRS90,RS91} leads to
reasonable results in the charm sector,  while it certainly fails to provide a
quantitative good description of the bottom baryons. This clearly indicates
that new explicit degrees of freedom have to be included in the effective
lagrangian in order to have the correct heavy quark limit.

To proceed it is important to observe that, to the lowest order in derivatives on
the Goldstone boson fields, Eq.~(\ref{Lck}) is the same as the lagrangian
Eq.~(\ref{lagrangian}) when only the heavy pseudoscalars are considered.
Furthermore, as argued in Refs.\cite{Nowak:1992um,Lee:1993jt,Lee:1993tg}$\!$,
as  the $h$ quark mass increases above the chiral scale $\Lambda_\chi$,
the Wess-Zumino term is expected to vanish, thereby turning off the last term of
(\ref{Lck}). Thus, the two lagrangians are indeed equivalent as far as the pseudoscalars
are concerned. However, as discussed in the previous sections, in order to have
the correct heavy quark limit one should explicitly take into account
the heavy vector degrees of freedom, which become degenerate with the pseudoscalars
as one approaches that limit. From an effective lagrangian point of view, the
vector mesons can be viewed as ``matter fields". There are several ways
of introducing vector matter fields.
Here we follow the hidden gauge symmetry (HGS) approach\cite{Bando:1987br}
in which case the non-anomalous effective lagrangian is
\begin{eqnarray}
{\cal L}_{0}&=&
-\frac{f_\pi^2}{4}{\,\mbox{Tr}}[{\cal D}_\mu\xi^{}_L\xi^\dagger_L
-{\cal D}_\mu\xi^{}_R\xi^\dagger_R]^2
-a\frac{f_\pi^2}{4}{\,\mbox{Tr}}[{\cal D}_\mu\xi^{}_L\xi^\dagger_L
+{\cal D}_\mu\xi^{}_R\xi^\dagger_R]^2
 -\frac12{\,\mbox{Tr}}(F_{\mu\nu}F^{\mu\nu}).
\nonumber \\
& &
\label{L0}
\end{eqnarray}
Here, ${\cal D}_\mu = \partial_\mu + ig_* U_\mu $ with
\begin{equation}
U_\mu=\frac12\left(
\begin{array}{cc}
\omega_\mu + \rho_\mu & \sqrt2 \Phi_\mu^{*\dagger} \\
\sqrt2 \Phi_\mu^*        &   \Psi^*_\mu\!
\end{array}\right) \ ,
\label{Umu}
\end{equation}
and $g_*$ is a gauge coupling constant to be specified later.
The field strength tensor of the vector mesons is $F_{\mu\nu}={\cal D}_\mu U_\nu-{\cal D}_\nu U_\mu$,
and the fields $\xi_L$ and $\xi_R$ are related to the chiral field
by $U(x) = \xi_L^\dagger \xi_R$.
The vector meson mass $M_{\rho,\omega}$ and the
$\rho\pi\pi$ coupling constant can be read off from the lagrangian,
\begin{eqnarray}
M_{\rho,\omega}^2 = ag_*^2f_\pi^2 \qquad ; \qquad g_{\rho\pi\pi}=\frac{a}{2} g_*.
\end{eqnarray}
The usual KSRF relation $m^2_\rho=2g_*^2f^2_\pi$,
and the universality of the vector-meson coupling
$g_{\rho\pi\pi}=g_*$,
can be used\cite{Bando:1987br} to fix the arbitrary parameter $a$ to 2.

The effective action should satisfy the same anomalous Ward identities
as does the underlying fundamental theory, QCD\cite{Wess:1971yu} .
In the presence of vector mesons $A^\mu_{L,R}$ associated with the external
(e.g. electroweak) gauge transformations, the general
form of the anomalous lagrangian is given by a special solution of the
anomaly equation plus general solutions of the homogeneous equation\cite{Fujiwara:1984mp}.
The former is the so-called gauged Wess-Zumino action $\Gamma^{g}_{WZ}$
(see e.g. Ref.\cite{Meissner:1987ge}
for details) and the latter, the anomaly free terms, can be made of
four independent blocks ${\cal L}_i$ whose explicit forms can be found
in Ref.\cite{Bando:1987br} .
Thus, for the anomalous processes we have
\begin{equation}
\Gamma_{an}=\Gamma^{g}_{WZ}[\xi^\dagger_L\xi^{}_R,A_L^{},A_R^{}]
+\sum_{i=1}^4 \ \gamma_i\ \int_{M_4}d^4x \ {\cal L}_i\ ,
\label{Gwz}
\end{equation}
with four arbitrary constants $\gamma_i$, which are determined by experimental
data. Vector meson dominance (VMD) in processes like
$\pi^0\rightarrow 2\gamma$ and $\gamma\rightarrow3\pi$ is very useful
in determining these constants.

As for the symmetry breaking one can take the form
\cite{Bramon:1994cb}
\begin{eqnarray}
{\cal L}_{sb} &=& - \frac{f_\pi^2}{4}{\,\mbox{Tr}}\left\{
({\cal D}_\mu\xi^{}_L\xi^\dagger_L-{\cal D}_\mu\xi^{}_R\xi^\dagger_R)^2
(\xi^{}_R\varepsilon_A^{}\xi^\dagger_L+\xi_L^{}\varepsilon_A^{}\xi^\dagger_R)
\right\} \nonumber \\
         & & -\frac{a f_\pi^2}{4}{\,\mbox{Tr}}\left\{
({\cal D}_\mu\xi^{}_L\xi^\dagger_L+{\cal D}_\mu\xi^{}_R\xi^\dagger_R)^2
(\xi^{}_R\varepsilon_V^{}\xi^\dagger_L+\xi_L^{}\varepsilon_V^{}\xi^\dagger_R)
\right\}\ .
\label{Lsb2}
\end{eqnarray}
The matrix $\varepsilon_{A(V )}$ is taken to be $\varepsilon_{A(V )} = \mbox{diag}(0, 0, c_{{A(V )}})$,
 where $c_{A(V)}$ are the SU(3)-breaking real parameters to be determined.
In terms of them one obtains
\begin{eqnarray}
m_{\Phi^*}^2 = \left( 1 + c_V \right) m_{\rho,\omega}^2
\qquad , \qquad
f^2_\Phi = \left( 1 + c_A \right)  f^2_\pi \ .
\label{SBm2}
\end{eqnarray}

Finally, we substitute the CK ansatz Eq.(\ref{Uck}),
(that is,  $\xi_L^\dagger=N_\pi\sqrt{U_\Phi}$ and
$\xi^{}_R=\sqrt{U_\Phi}N_\pi$)
into the total effective action
\begin{equation}
{\Gamma}={\Gamma}_0 + {\Gamma}_{an}+ {\Gamma}_{sb}\ ,
\end{equation}
where ${\Gamma}_0$ and ${\Gamma}_{sb}$ are obtained from
the lagrangians
Eq.~(\ref{L0}) and Eq.~(\ref{Lsb2}), respectively, and
the action ${\Gamma}_{an}$
is given in Eq.~(\ref{Gwz}).
One may  check that the resulting lagrangian contains all the terms of
Eq.~(\ref{lagrangian}). Explicitly, one gets\cite{Min:1994qq}
\begin{eqnarray}
{\cal L} &=& {\cal L}^{Sk}_l + D_\mu \Phi D_\mu \Phi^\dagger
- m_\Phi^2 \Phi \Phi^\dagger
-\frac12 \Phi^{*\mu\nu} \Phi^{*\dagger}_{\mu\nu}
+ m^2_{\Phi^*} \Phi^{*\mu} \Phi^{*\dagger}_\mu \nonumber \\
& & -\sqrt2 m_{\Phi^*}^{} (\Phi A^\mu \Phi^{*\dagger}_\mu +
\Phi^*_\mu A^\mu \Phi^\dagger)
+\frac{i}{2} c_4 g_*^2 \varepsilon^{\mu\nu\lambda\rho} (\Phi^*_{\mu\nu}A_\lambda
\Phi^{*\dagger}_\rho
+ \Phi^*_\lambda A_\rho \Phi^{*\dagger}_{\mu\nu}) + \dots \ , \nonumber
\\
& &
\label{Lhml}
\end{eqnarray}
where the light vector meson fields $\rho_\mu$ and
$\omega_\mu$ have been replaced
by $2i\ \! V_\mu/g_*$ and $(c_1-c_2)i6 \pi^2 B_\mu/g_*f_\pi^2$,
respectively,
and terms with higher derivatives acting on the pion fields have not
been explicitely written.
Comparing Eq.~(\ref{Lhml}) with Eq.~(\ref{lagrangian}), we obtain
two relations
\begin{equation}
f_{_Q}=-\sqrt2m_{\Phi^*}, \qquad \mbox{and} \qquad g_{_Q}= i\gamma_4g_*^2\ .
\end{equation}
The first relation implies that
\begin{equation}
\frac{f_{_Q}}{2m_{\Phi^*}}=-\frac{1}{\sqrt2}\ ,
\end{equation}
which is quite close to the expected heavy quark limit result Eq.(\ref{hqr}) with
$g=-0.75$ evaluated with the NRQM in Sec.~2. Using this relation
and assuming that the VMD works in the heavy meson sector,
in which case $\gamma_4=iN_c/16\pi^2$, one obtains $g_*$
in the heavy quark limit, i.e.
\begin{equation}
g_*=\sqrt{\frac{16\pi^2}{\sqrt2 N_c}}\simeq 6 \hskip 1.5cm
(\mbox{with $N_c$=3})\ .
\end{equation}
which is close to $g_* = g_{\rho\pi\pi}$ found in the light sector.
These results seem to indicate that, in principle, it might be possible to
construct an effective soliton model which could be used to describe both the
strange sector and the heavier sectors. Of course, further work is definitely
required in order to test in detail the feasibility of this ambitious
program.

To conclude this section, we note that there is an alternative method\cite{Yabu:1987hm}
to describe strange hyperons within topological soliton models (for reviews
see e.g. Ref.\cite{Weigel:1995cz}). That method is based on treating strange degrees
of freedom as light and, thus, on the introduction of rotational $SU(3)$ collective
quantization. It is clear that this treatment becomes better the closer one is
to the limit $m_K \rightarrow 0$. It has been suggested\cite{Kaplan:1989fc},
however, that even in such a limit the bound state picture is applicable.
In the present context this brings in the very interesting question
concerning the possibility of having a unified framework that may
allow to smoothly interpolate between the chiral symmetry limit
and the heavy quark limit.

\section{Summary and conclusions}
\label{summary}

Heavy baryons represent an extremely interesting problem since they combine the dynamics of the heavy
and light sectors of the strong interactions. In this contribution we have reviewed the work done on the
description of heavy baryons as heavy-meson--soliton bound systems. We have first discussed how these
bound systems can be obtained in the infinite heavy quark limit using effective lagrangians that respect
both chiral symmetry and heavy quark symmetry. Next, we have shown how the effects due to finite heavy
quark masses can be accounted for, and compared the resulting heavy baryon spectra with existing
quark model and empirical results. This comparison indicates that, even though room for improvement is certainly
left, the bound heavy-meson--soliton models are reasonably successful in reproducing those results.
Finally, we have addressed some issues related to a possible connection between the usual bound state approach
to strange hyperons and that for heavier baryons. We have shown that there are some indications that it might be
possible to construct an effective soliton model which could be used to describe baryons formed by quarks of any flavor.
Of course, further work is definitely required in order to test in detail the feasibility of this ambitious program.
We finish by recalling that, although in recent years there has been an enormous progress in both the theoretical
and experimental aspects of the heavy baryon physics, many problems still remain to be resolved. For example,
most of the $J^P$ quantum numbers of the heavy baryons have not been yet determined experimentally, but are assigned
on the basis of quark model predictions. In this sense, the insight obtained from alternative models
such as the bound state soliton model discussed in the present contribution might be particularly useful.

\section*{Acknowledgements}

I would like to thank J.L. Goity, B-Y. Park, M. Rho and
D.O. Riska for useful comments.
This work was supported in part by CONICET (Argentina) grant \# PIP 6084 and
by ANPCyT  (Argentina) grants \# PICT04 03-25374 and \# PICT07 03-00818.

\end{document}